\documentclass[aps,epjc,twocolumn,floatfix,superscriptaddress,nofootinbib,showpacs,showkeys]{revtex4}
\usepackage{graphicx}
\usepackage{dcolumn}  
\usepackage{bm}       
\usepackage{palatino}
\usepackage{epsfig}
\begin{document}
\title{Average formation length in string model
}
\author{L.~Grigoryan \\
Yerevan Physics Institute, Br.Alikhanian 2, 375036 Yerevan, Armenia }
\begin {abstract}
\hspace*{1em}
The space-time scales of the hadronization process in the framework of string
model are investigated. It is shown that the average formation lengths of
pseudoscalar mesons, produced in semi-inclusive deep inelastic scattering (DIS)
of leptons on different targets, depend from their electrical charges.
In particular the average formation lengths of positively charged hadrons are
larger than of negatively charged ones. This statement is fulfiled for all using
scaling functions, for $z$ (the fraction of the virtual photon energy
transferred to the detected hadron) larger than $0.15$, for
all nuclear targets and any value of the Bjorken scaling variable $x_{Bj}$.
In all cases, the main mechanism is direct production of pseudoscalar mesons.
Including in consideration additional mechanism of production in result of decay of
resonances, leads to decrease of average formation lengths.
It is shown that the average formation lengths of positively
(negatively) charged mesons are slowly rising (decreasing) functions of $x_{Bj}$.
The obtained results can be important, in particular, for the understanding of
the hadronization process in nuclear environment.
\end {abstract}
\maketitle
\normalsize
\hspace*{1em}
Hadronization is a process that leads from partons produced in an initial
hard interaction to the final hadrons observed experimentally.
Two aspects of hadronization: (i) the spectra of hadrons produced and
their kinematical dependences;
(ii) the space-time evolution of the process, at present
investigated not equally well. First
of them has been studied extensively in $e^{+}e^{-}$ annihilation and lepton-nucleon
DIS both experimentally and theoretically. As a result the spectra of hadrons and
their kinematical dependences are rather well known. There are a few successful
theoretical models, which give transition from initial partons to final hadrons
through sets of fragmentation functions. The second aspect of hadronization, the
space-time evolution of the process, despite on its importance, was investigated
relatively little. Although the string model allows to study the space-time scale
of the hadronization process, only a few works were performed in this
direction~\cite{chmaj,Bi_Gyul}.
In~\cite{Bi_Gyul} it was claimed, that for hadrons (as composite particles) the
very notion of formation length is ambiguous because different constituents of a
hadron can be produced at different lengths. It is then an open and model-dependent
question which of the two length scales plays more important role in hadronization
process: (i) constituent formation length $l_c$ which is the distance between DIS
point and the point where first constituent parton of the final hadron arises; (ii)
the yo-yo formation length $l_y$ which is the distance between DIS point and point
where two constituent partons of the final hadron meet first time and form particle
with quantum numbers of final hadron but without its "sea". It is worth to note
that these two length scales connected by simple way.\\
\hspace*{1em}
The string model can give information about space-time scale of hadronization, but
it gives nothing about development of hadronic properties of string during
hadronization process. At present it is supposed, that the unique way to get
information about latter is the experimental and phenomenological study of
hadronization process in atomic nuclei, where string interacts with nuclear medium.
It is assumed that constituent formation length plays more important role in the
hadronization process, because beginning with this scale the piece of string with
a constituent parton on the slow end interacts with hadronic cross section. As was
pointed out in Ref.~\cite{Bi_Gyul} this result follows from the comparison with the
data on fragmentation of $30 GeV$ pions into $p$ and $\bar{p}$~\cite{abreu}.
In~\cite{akopov} the formation lengths of pions were presented in form
$l = (1 - w)l_{c} + wl_{y}$, where $w$ is the probability that formation length is
$l_{y}$. Comparison with experimental data~\cite{airap2} showed, in case when $l_{c}$
and $l_{y}$ were calculated in the framework of standard Lund model, that
$w = 0.1 - 0.17$.
This result confirmed conclusion of Ref.~\cite{Bi_Gyul} about importance of
constituent formation length. Further we will consider it as a formation length. It is
a function of variables $\nu$ and $z$ (the energy of virtual photon and the fraction
of this energy carried away by the final hadron with energy $E_h$ ($z=E_h/\nu$))
and can change, as we will see below, in wide region from zero to tens (may be even
hundreds) femtometers.\\
\hspace*{1em}
In Refs.~\cite{akopov2,akopov3,grig} it was shown that a ratio of multiplicities
for the nucleus and deuterium can be presented in the form of a function of a
single variable which has the physical meaning of the formation length (time)
of the hadron. This scaling was verified for the case of charged pions by
HERMES experiment~\cite{airap3}. Now HERMES experiment prepares two-dimensional
analysis of nuclear attenuation data using more precise definition of the
formation length of hadron presented in this work.\\
\hspace*{1em}
In the string model, for the construction of fragmentation functions, the scaling
function $f(z)$ is introduced (see, for instance, Refs.~\cite{andersson,field,lund}). 
It is defined by the condition that $f(z)dz$ is the probability that the first
hierarchy (rank 1) primary meson carries away the fraction of energy $z$ of the
initial string. We use three different scaling functions for calculations:\\
(i) standard Lund scaling function~\cite{lund}
\begin{eqnarray}
f(z) = (1 + C)(1 - z)^{C} ,
\end{eqnarray}
where $C$ is the parameter which controls the steepness of the standard Lund
fragmentation function;\\
(ii) symmetric Lund scaling function~\cite{lund,seostrand}
\begin{eqnarray}
f(z) = Nz^{-1}(1 - z)^{a}exp(-bm_{\perp}^{2}/z) ,
\end{eqnarray}
where $a$ and $b$ are parameters of model, $m_{\perp}=\sqrt{m_{h}^{2}+p_{\perp}^{2}}$
is the transverse mass of final hadron, $N$ is normalization factor;\\
(iii) Field-Feynman scaling function~\cite{field}:
\begin{eqnarray}
f(z) = 1 - a + 3a(1 - z)^{2} ,
\end{eqnarray}
where a is parameter of model. We will specify parameters below, when will discuss
the details of calculations.\\
\hspace*{1em}
In the further study we will use the average value of the formation length defined
as $L_{c}^{h} = <l_c>$.\\
\hspace*{1em}
The consideration is convenient to begin from $L_{c}^{h}$ direct, $L_{c}^{h(dir)}$,
which takes into account the direct production of hadrons:
\begin{eqnarray}
{L_{c}^{h(dir)} = \int_0^{\infty} ldlD_{c}^{h}(L,z,l)/\int_0^{\infty} 
dlD_{c}^{h}(L,z,l)}
\hspace{0.3cm},
\end{eqnarray}
where $L = \nu/\kappa$ is the full hadronization length, $\kappa$ is the string tension
(string constant), $D_{c}^{h}(L, z, l)$ is the distribution of the constituent formation
length $l$ of hadrons carrying fractional energy $z$.
\begin{eqnarray}
\nonumber
D_{c}^{h}(L,z,l) = \Big(C_{p1}^{h}f(z)\delta(l-L+zL)+
\end{eqnarray}
\begin{eqnarray}
C_{p2}^{h}\sum_{i=2}^{n}D_{ci}^{h}(L,z,l)\Big)\theta(l)\theta(L-zL-l)
\hspace{0.3cm}.
\end{eqnarray}
The functions $C_{p1}^{h}$ and $C_{p2}^{h}$ are the probabilities that in
electroproduction process on proton target the valence quark compositions for leading
(rank 1) and subleading (rank 2) hadrons will be obtained. Similar functions were
obtained in~\cite{akopov1} for more general case of nuclear targets.
In eq.(5) $\delta$- and $\theta$-functions arise as a consequence of energy conservation
law. The functions $D_{ci}^{h}(L,z,l)$ are distributions of the constituent formation
length $l$ of the rank $i$ hadrons carrying fractional energy $z$. For calculation of
distribution functions we used recursion equation from Ref.~\cite{Bi_Gyul}.\\
\hspace*{1em}
The simple form of $f(z)$ for standard Lund model (eq.(1)) allows to sum the sequence of
produced hadrons over all ranks ($n=\infty$). The analytic expression for the
distribution function in this case is:
\begin{eqnarray}
\nonumber
D_{c}^{h}(L,z,l) = L(1+C)\frac{l^C}{(l+zL)^{C+1}}\times
\end{eqnarray}
\begin{eqnarray}
\Big(C_{p1}^{h}\delta(l-L+zL)+C_{p2}^{h}\frac{1+C}{l+zL}\Big)\theta(l)\theta(L-zL-l) 
\hspace{0.3cm}.
\end{eqnarray}
In~\cite{chmaj,Bi_Gyul} (see also~\cite{czyzewski}) eq.(6) was obtained for the special
case $C_{p1}^{h} = C_{p2}^{h} = 1$. Another special case was considered in 
Ref.~\cite{accardi}. It was supposed, that in electroproduction process only the $u$
quarks are knocked out, which then turn into observed hadrons. It is clear, that this 
approximation is more or less valid for proton target in region of large enough values
of Bjorken scaling variable. For positively charged mesons it does not differ from the
first special case, for negatively charged ones it corresponds to special case 
$C_{p1}^{h} = 0, C_{p2}^{h} = 1$. Nevertheless, in this rough approximation it was obtained,
that the average formation lengths for positively charged hadrons are
larger than for negatively charged ones.\\
\hspace*{1em}
Unfortunately, in case of more complicated scaling functions presented in
eqs.(2) and (3) the analytic summation of the sequence of produced hadrons over all
ranks is impossible. In these cases we limited ourself by $n=10$ in eq.(5).\\
\hspace*{1em}
As it is well known, essential contributions in the spectra of pions and kaons
from the decays of vector mesons are expected. But so far, the formation lengths of
pseudoscalar mesons were considered without taking into account this possibility.\\
\hspace*{1em}
We build the distributions for daughter mesons from the ones for parent resonances
followed the method used in ~\cite{andersson,field} for the construction of fragmentation 
functions for daughter mesons.\\
\hspace*{1em}
The distribution function of the constituent formation length $l$ of the daughter
hadron $h$ which arises in result of decay of parent resonance $R$ and carries away the 
fractional energy $z$ is denoted $D_c^{R/h}(L, z, l)$. It can be computed from the convolution
integral:
\begin{eqnarray}
\nonumber
D_c^{R/h}(L, z, l) = d^{R/h}
\int_{z_{down}^{R/h}}^{z_{up}^{R/h}} \frac{dz'}{z'}D_{c}^{R}(L,z',l)\times
\end{eqnarray}
\begin{eqnarray}
f^{R/h}\Big(\frac{z}{z'}\Big) 
\hspace{0.3cm},
\end{eqnarray}
where $z_{up}^{R/h} = min(1,z/z_{min}^{R/h})$ and $z_{down}^{R/h} = min(1,z/z_{max}^{R/h})$,
$z_{max}^{R/h}$ ($z_{min}^{R/h}$) is maximal (minimal) fraction of the energy of parent 
resonance, which can be carried away by the daughter meson.\\
\hspace*{1em}
Let us consider the two-body isotropic decay of resonance $R$, $R \to h_{1}h_{2}$, and
denote the energy and momentum of the daughter hadron $h$ ($h = h_{1}$ or $h_{2}$), in
the rest system of resonance, $E_{h}^{(0)}$ and $p_{h}^{(0)}$, respectively.
In the coordinate system where resonance has energy and momentum equal $E_R$ and $p_R$,
\begin{eqnarray}
z_{max}^{R/h} = \frac{1}{m_{R}}\Big(E_{h}^{(0)} +
\frac{p_R}{E_R}p_{h}^{(0)}\Big)
\hspace{0.3cm},
\end{eqnarray} 
\begin{eqnarray}
z_{min}^{R/h} = \frac{1}{m_{R}}\Big(E_{h}^{(0)} -
\frac{p_R}{E_R}p_{h}^{(0)}\Big)
\hspace{0.3cm},
\end{eqnarray}
where $m_R$ is the mass of resonance $R$. In the laboratory (fixed target) system the
resonance usually fastly moves, i.e. $p_R/E_R \to 1$.\\
\hspace*{1em}
The constants $d^{R/h}$ can be found from the branching ratios in the decay process
$R \to h$. We will present their values for interesting for us cases below.\\
\hspace*{1em}
The distributions $f^{R/h}(z)$ are determined from the decay process of the resonance
R, with momentum $p$ into the hadron $h$ with momentum $zp$. We assume that the momentum
$p$ is much larger than the masses and the transverse momenta involved.\\
\hspace*{1em}
In analogy with eq.(4) we can write the expression for the average value of the
formation length $L_{c}^{R/h}$ for the daughter meson $h$ produced in result of decay
of the parent resonance $R$ in form:
\begin{eqnarray}
{L_{c}^{R/h} = \int_0^{\infty} ldlD_{c}^{R/h}(L,z,l)/
\int_0^{\infty} dlD_{c}^{R/h}(L,z,l)}
\hspace{0.3cm}.
\end{eqnarray}\\
Here it is need to give some explanations. We can formally consider $L_{c}^{R/h}$ as the
formation length of daughter meson $h$ for two reasons: (i) the parent resonance and
daughter hadron are the hadrons of the same rank~\cite{field}, which have common
constituent quark; (ii) according to above discussions, beginning from this distance the
chain consisting from prehadron, resonance and final meson $h$ interacts (in nuclear
medium) with hadronic cross sections.\\
\hspace*{1em}
The general formula for $L_{c}^{h}$ for the case when a few resonances
contribute can be written in form:
\begin{eqnarray}
\nonumber
L_{c}^{h} = \int_0^{\infty} ldl\Big(\alpha_{ps}D_c^{h}(L,z,l) + 
\alpha_{v}\sum_{R}D_c^{R/h}(L,z,l)\Big)/
\end{eqnarray}
\begin{eqnarray}
\int_0^{\infty} dl\Big(\alpha_{ps}D_c^{h}(L,z,l) +             
\alpha_{v}\sum_{R}D_c^{R/h}(L,z,l)\Big)
\hspace{0.15cm},
\end{eqnarray}
where $\alpha_{ps}$ ($\alpha_{v}$) is the probability that $q\bar{q}$ pair
turns into pseudoscalar (vector) meson.
Following Refs.~\cite{lund,field} we use condition
$\alpha_{ps} = \alpha_{v} = \frac{1}{2}$.\\
\hspace*{1em}
Let us now discuss the details of model, which are necessary for
calculations. We will cosider four kinds of pseudoscalar mesons $\pi^{+}$,
$\pi^{-}$, $K^{+}$ and $K^{-}$ electroproduced on proton, neutron and nuclear
targets. The scaling function $f(z)$ in eq.(1) has single free parameter $C$.
It is known~\cite{lund} that comparison with experimental data gives limitation
on its possible values $C = 0.3 - 0.5$. Our calculations showed that changing
of parameter $C$ from the minimal value $C = 0.3$ to the maximal value
$C = 0.5$ leads to the small increasing of the average formation
lengths. Therefore further we will present
results for standard Lund model obtained for $C = 0.3$ only.
For symmetric Lund scaling function we use parameters~\cite{seostrand}
$a=0.3$, $b=0.58 GeV^{-2}$, and for Field-Feynman scaling function~\cite{field} the
value $a=0.77$ for the single parameter $a$.
Next parameter, which is necessary for the calculations in the
framework of string model is the string tension. It was fixed at a static value
determined by the Regge trajectory slope~\cite{kappa,seostrand}
\begin{eqnarray}
\kappa = 1/(2\pi\alpha'_R) = 1 GeV/fm \hspace{0.3cm}.
\end{eqnarray}
\hspace*{1em}
Now let us turn to the functions $C_{p1}^{h}$ and $C_{p2}^{h}$.
For pseudoscalar mesons they have form.:
\begin{eqnarray}
\nonumber
C_{p1}^{\pi^{+}} = \frac{\frac{4}{9}u(x_{Bj},Q^2) + \frac{1}{9}\bar{d}(x_{Bj},Q^2)}
{\sum_{q=u,d,s} e_{q}^{2}(q(x_{Bj},Q^2) + \bar{q}(x_{Bj},Q^2))} \gamma_{q} ,
\end{eqnarray}
\begin{eqnarray}
\nonumber
C_{p1}^{\pi^{-}} = \frac{\frac{4}{9}\bar{u}(x_{Bj},Q^2) + \frac{1}{9}d(x_{Bj},Q^2)}
{\sum_{q=u,d,s} e_{q}^{2}(q(x_{Bj},Q^2) + \bar{q}(x_{Bj},Q^2))} \gamma_{q} ,
\end{eqnarray}
\begin{eqnarray}
\nonumber
C_{p2}^{\pi^{+}} = C_{p2}^{\pi^{-}} = \gamma_{q}^{2} ,
\end{eqnarray}
\begin{eqnarray}
\nonumber
C_{p1}^{K^{+}} = \frac{\frac{4}{9}u(x_{Bj},Q^2)\gamma_{s} + 
\frac{1}{9}\bar{s}(x_{Bj},Q^2)\gamma_{q}}
{\sum_{q=u,d,s} e_{q}^{2}(q(x_{Bj},Q^2) + \bar{q}(x_{Bj},Q^2))}  ,
\end{eqnarray}
\begin{eqnarray}
\nonumber
C_{p1}^{K^{-}} = \frac{\frac{4}{9}\bar{u}(x_{Bj},Q^2)\gamma_{s} + 
\frac{1}{9}s(x_{Bj},Q^2)\gamma_{q}}
{\sum_{q=u,d,s} e_{q}^{2}(q(x_{Bj},Q^2) + \bar{q}(x_{Bj},Q^2))}  ,
\end{eqnarray}
\begin{eqnarray}
\nonumber
C_{p2}^{K^{+}} = C_{p2}^{K^{-}} = \gamma_{q} \gamma_{s} ,
\end{eqnarray}
where $x_{Bj} = \frac{Q^{2}}{2m_{p}\nu}$ is the Bjorken's scaling variable;
$Q^{2} = -q^{2}$, where $q$ is the 4-momentum of virtual photon; $m_{p}$ is
proton mass; $q(x_{Bj},Q^2) (\bar{q}(x_{Bj},Q^2))$, where $q=u,d,s$ are quark
(antiquark) distribution functions for proton. Easily to see, that functions
$C_{pn}^{h}$ for hadrons of higher rank ($n > 2$) coincide with ones for
second rank hadron $C_{pn}^{h} \equiv C_{p2}^{h}$. This fact was already used
for construction of eq.(5). For neutron and nuclear targets more general
functions $C_{fi}^{h} (i = 1,2)$ from~\cite{akopov1} are used. 
Functions for resonances can be built in analogy with the above equations.
For example, $C_{pi}^{\rho^{+}} = C_{pi}^{\pi^{+}}$, where $i=1,2$.\\
\hspace*{1em}
All calculations were performed at fixed value of $\nu$ equal $10 GeV$.
Calculations of $z$ - dependence were performed at fixed value of
$Q^{2}$ equal $2.5 GeV^{2}$, which give $x_{Bj} \approx 0.133$. For quark (antiquark)
distributions in proton the parameterization for leading order parton distribution
functions from~\cite{grv} was used. We assume, that new $q\bar{q}$ pairs are
$u\bar{u}$ with probability $\gamma_{u}$, $d\bar{d}$ with probability $\gamma_{d}$
and $s\bar{s}$ with probability $\gamma_{s}$. It is followed from isospin symmetry
that $\gamma_{u} = \gamma_{d} = \gamma_{q}$.
We use two sets of values for $\gamma$: for Lund model~\cite{seostrand}
$\gamma_{u} : \gamma_{d} : \gamma_{s} = 1 : 1 : 0.3$; and for Field-Feynman
model $\gamma_{u} : \gamma_{d} : \gamma_{s} = 1 : 1 : 0.5$.\\
\hspace*{1em}
We take into account that part of pseudoscalar mesons can be produced from decays
of resonances. As possible sources of $\pi^{+}$, $\pi^{-}$, $K^{+}$ and $K^{-}$
mesons we consider $\rho^+$, $\rho^0$, $\omega$, $K^{*+}$ and $\bar{K}^{*0}$;
$\rho^-$, $\rho^0$, $\omega$, $K^{*-}$ and $K^{*0}$; $K^{*+}$, $K^{*0}$ and $\phi$;
$K^{*-}$, $\bar{K}^{*0}$ and $\phi$ mesons, respectively.
The contributions of other resonances are neglected.\\
\begin{figure}[!htb]
\begin{center}
\epsfxsize=8.cm
\epsfbox{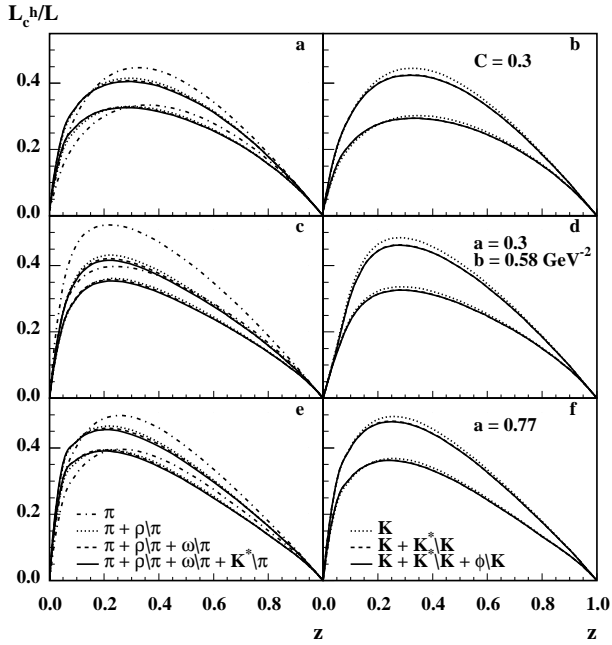}
\end{center}
\caption{\label{xx1}
{\it Average formation lengths for electroproduction of hadrons on proton,
normalized on $L$, are presented as a function of $z$. The contributions of direct
hadrons as well as of the sum of direct and produced from decay of resonances
hadrons are presented. Upper curves represent formation lengths of positively
charged hadrons and lower curves of negatively charged ones. 
Results for standard Lund model are presented on panels a, b; for symmetric
Lund model on panels c, d; for Field-Feynman model on panels e, f, respectively. 
}}
\end{figure}
\hspace*{1em}
In Ref.~\cite{andersson} were presented simple expressions for $f^{R/h}$ which
are close enough to the experimental data. In case of pions we interested in
$f^{\rho/\pi}$, $f^{\omega/\pi}$ and $f^{K^{*}/\pi}$, in case of kaons
in $f^{K^{*}/K}$ and $f^{\phi/K}$. (Sometimes we omit the charge symbols in our
expressions, but everywhere it is imply that the different necessary charge
states for parent and daughter hadrons are taken into account.) The function
$f^{\omega/\pi}$ has form $f^{\omega/\pi}(z) = 2(1 - z)$. For the other functions
common expression $f^{R/h}(z) = 1/(z_{max}^{R/h}-z_{min}^{R/h})$ is used. 
The values of $z_{max}^{R/h}$ and $z_{min}^{R/h}$ it is easily to obtain from
eqs.(8) and (9). For instance, for the $\rho$ meson decay into pions we receive 
$z_{max}^{\rho/\pi} \approx 0.965$ and $z_{min}^{\rho/\pi} \approx 0.035$.\\
\hspace*{1em}
For $\pi^{+}$ and $\pi^{-}$ mesons we have $d^{\rho/\pi} = \frac{1}{2}$,
$d^{\omega/\pi} = 0.3$ and $d^{K^{*}/\pi} = \frac{1}{3}$. 
For $K^{+}$ and $K^{-}$ mesons we have $d^{K^{*+}/K^{+}} = d^{K^{*-}/K^{-}} = \frac{1}{6}$;
$d^{K^{*0}/K^{+}} = d^{\bar{K}^{*0}/K^{-}} = \frac{1}{3}$; $d^{\phi/K} \approx \frac{1}{4}$.\\
\hspace*{1em}
In Fig.1 the average formation lengths for electroproduction of different
pseudoscalar mesons on proton,
normalized on $L$, are presented as a function of $z$. 
The formation lengths for pions (panels a, c and e) and kaons (b, d, f) are presented.
The contributions of direct
hadrons as well as of the sum of direct and produced from decay of resonances
hadrons are presented. Upper curves represent formation lengths of positively
charged hadrons and lower curves of negatively charged ones.
Results for standard Lund model are presented on panels a, b; for symmetric
Lund model on panels c, d; for Field-Feynman model on panels e, f, respectively.
The values of parameters using in calculations are presented also.
Of course results of different models are quantitatively differ, but qualitatively
they have the same behavior as functions of $z$. Therefore further, for illustration,
we will use the results of symmetric Lund model only.\\
\hspace*{1em}
Let us briefly discuss why the average formation lengths of positively charged hadrons are
larger than of negatively charged ones. It happens due to the large probability to knock
out $u$ quark in result of DIS (even in case of neutron target). The knocked out quark 
enter
in the composition of leading hadron, which has maximal formation length.
The $K^{+}$ meson has average formation length larger than $\pi^{+}$ meson, because
in first case the influence of resonances is smaller. The $K^{-}$ meson has average
formation length smaller than $\pi^{-}$ meson, because it is constructed from "sea"
quarks and practically can not be leading hadron, whereas $\pi^{-}$ meson can be leading
hadron due to $d$ quark entering in its composition.
\begin{figure}[!htb]
\begin{center}
\epsfxsize=8.cm
\epsfbox{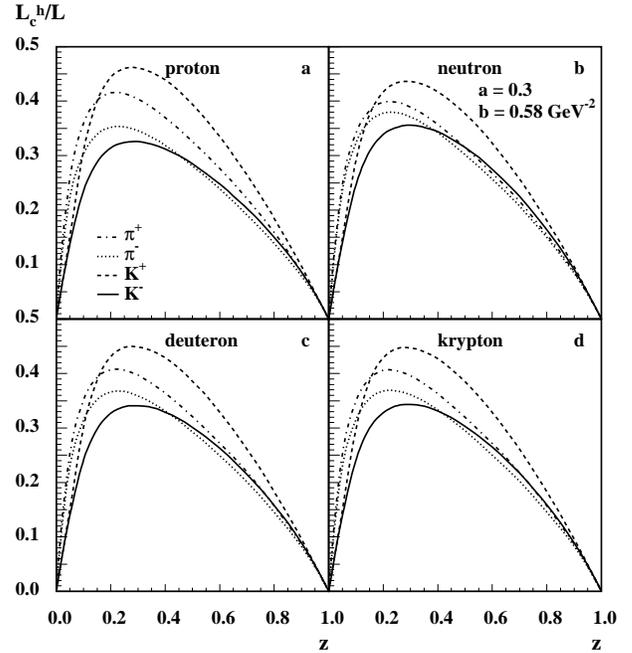}
\end{center} 
\caption{\label{xx2}
{\it Average formation lengths for electroproduction of $\pi^+$, $\pi^-$, $K^+$
and $K^-$  mesons on different targets in symmetric Lund model, normalized on $L$,
as a function of $z$.
}}
\end{figure}
\hspace*{1em}
In Fig.2 the average formation lengths for electroproduction of $\pi^+$, $\pi^-$,
$K^+$ and $K^-$  mesons on different targets in symmetric Lund model, normalized on $L$,
as a function of $z$ are presented. It is taken into account, that mesons can be
produced directly or from decay of resonances. Easily to see, that positively
charged hadrons have larger $L_{c}^{h}$ than negatively charged ones. This statement
is right even for neutron target. At large enough $z$ (for instance $z > 0.2$)
$K^+$ meson has maximal average formation length and $K^-$ meson minimal. Difference
between $L_{c}^{h}$ of positively and negatively charged hadrons is maximal on
proton target and minimal on neutron one.
It is worth to note, that results for deuteron coincide, in our approach,
with results for any nuclei with $Z = N$, where $Z$ ($N$) is number of protons (neutrons).
Average formation lengths of hadrons on krypton nucleus, which has essential excess of
neutrons, do not differ considerably from the ones on nuclei with $Z = N$.
\begin{figure}[!htb]
\begin{center}  
\epsfxsize=8.cm
\epsfbox{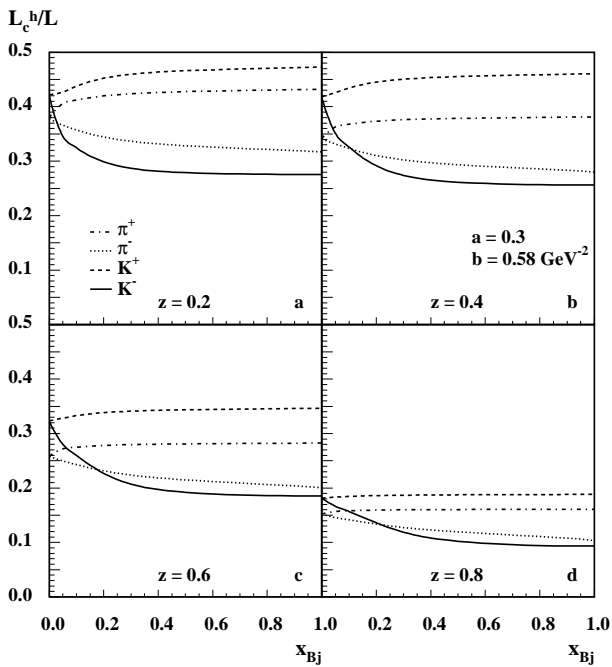}
\end{center}  
\caption{\label{xx3}
{\it Average formation lengths for electroproduction of $\pi^+$, $\pi^-$, $K^+$ and
$K^-$ mesons on proton target in symmetric Lund model, normalized on $L$, as a function
of $x_{Bj}$.
}}
\end{figure}
\hspace*{1em}
In Fig.3 the average formation lengths for electroproduction of $\pi^+$, $\pi^-$,
$K^+$ and $K^-$  mesons on proton in symmetric Lund model, normalized on $L$, as a
function of $x_{Bj}$ are presented. It is taken into account, that mesons can be
produced directly or from decay of resonances.\\
\hspace*{1em}
We obtained, for the first time, the average formation lengths for different
pseudoscalar mesons in the eletroproduction process on proton, neutron, deuteron and
krypton.
Main conclusions are: (i) positively charged pseudoscalar mesons ($\pi^{+}$ and $K^{+}$)
have the formation lengths larger than negatively charged ones ($\pi^{-}$ and $K^{-}$) on
all targets; (ii) contribution from the decay of resonances is maximal for $\pi^{+}$
mesons (reach $\sim 20\%$ in case of symmetric Lund model),
for $\pi^{-}$ and $K^{+}$ mesons it reaches a few percents,
the formation length of $K^{-}$ mesons practically does not feel contribution from
resonances; as it was expected, in case of pions maximal contribution gives $\rho$ meson,
in case of kaons $K^{*}$ meson.\\
It is worth to note that in string
model the formation length of the leading (rank 1) hadron $l_{c1} = (1 - z)\nu/\kappa$
does not depend from type of process, kinds of hadron and target.
We want to stress, that obtained result depends from the type
of process, kinds of targets and observed hadrons mainly due to presence of higher rank
hadrons. Including in consideration hadrons
produced from decay of resonances diminishes constituent formation length. It happens
because for producing of hadron with fractional energy $z$ we must have resonance with
$z^{'}$ larger than $z$. The larger is the fractional energy, the shorter is the
formation length.
Of course this statement is right for the large enough $z$ (for instance $z > 0.2$).\\
\hspace*{1em}
Calculations performed with different scaling functions: standard Lund~\cite{lund},
Field-Feynman~\cite{field} and symmetric Lund~\cite{lund}
showed, that although the numerical values of
average formation lengths slightly shift, qualitatively they have the same behavior
(see Fig.1).\\ 
\hspace*{1em}
Which sizes can reach the average formation length?
At fixed $x_{Bj}$ it is proportional to $\nu$. Consequently it will rise with $\nu$ and can
reach sizes much larger than nuclear sizes at very high energies. Naturally, this
conclusion is true if the formalism of string model will continue to work at such high
energies.\\
\hspace*{1em}
At present the hadronization in nuclear medium is widely studied both experimentally and
theoretically. It is well known, that there is nuclear attenuation of final hadrons.
Unfortunately it does not clear, which is the true mechanism of such attenuation: (i)
final state interactions of prehadrons and hadrons in nucleus (absorption mechanism); or
(ii) gluon bremmstrahlung of partons (produced in DIS) in nuclear medium, whereas
hadronization takes place far beyond nucleus (energy loss mechanism). We hope, that
results obtained in this letter can be useful for the understanding of this problem.\\
\vspace{0.3cm}\\
{\bf Acknowledgements}
\vspace{0.3cm}\\
\hspace*{1em}
I am grateful for stimulating discussions to N.Akopov, H.P.Blok, G.Elbakian and
I.Lehmann who read the paper and made many useful comments.

\end{document}